\documentclass[12pt,preprint]{aastex}

\slugcomment{To appear in ApJ}

\shorttitle{H53$\alpha$ Observations of the Orion BN/KL Region}
\shortauthors{Rodr\'\i guez et al.}

\begin{document}


\title{Resolving the Structure and Kinematics of the BN Object \\ at $0\rlap.{''}2$
Resolution}


\author{Luis F. Rodr\'\i guez}
\affil{Centro de Radioastronom\'\i a y Astrof\'\i sica, UNAM, Morelia 58089, M\'exico}
\email{l.rodriguez@astrosmo.unam.mx}

\author{Luis A. Zapata\altaffilmark{1}}
\affil{Max-Planck-Institut f\"ur Radioastronomie, Auf dem H\"ugel 69, 53121 Bonn, Germany}
\email{lzapata@mpifr-bonn.mpg.de}

\and

\author{Paul T. P. Ho\altaffilmark{1}}
\affil{Academia Sinica Institute of Astronomy and Astrophysics, Taipei 106, Taiwan}
\email{pho@asiaa.sinica.edu.tw}


\altaffiltext{1}{Harvard-Smithsonian Center for Astrophysics,
and 
Submillimeter Array, 60 Garden Street, Cambridge, MA 02138, USA}


\begin{abstract}
We present sensitive 7 mm observations of the H53$\alpha$ recombination line
and adjacent continuum, made toward the Orion BN/KL region. In the continuum
we detect the BN object, the radio source 
I (GMR I) and the radio counterpart of the infrared source n 
(Orion-n). Comparing with observations made at similar angular resolutions
but lower frequency, we discuss the spectral indices and angular sizes of these
sources. In the H53$\alpha$ line we only detect the BN object.
This is the first time that radio recombination lines are detected from this source.
The LSR radial velocity of BN from the H53$\alpha$ line, $v_{LSR} = 20.1 \pm 2.1$
km s$^{-1}$,
is consistent with that found from previous studies in near-infared lines.
While the continuum emission is expected to have
considerable optical depth at 7 mm, the observed H53$\alpha$ line emission is consistent with 
an optically-thin nature and we discuss possible explanations
for this apparent discrepancy.
There is evidence of a velocity gradient, with the NE part of BN being redshifted by $\sim
10$ km s$^{-1}$ with respect to the SW part. This is consistent with the
suggestion of Jiang et al. that BN may be driving an ionized outflow
along that direction. 
\end{abstract}





\section{Introduction}

Located at about 1$'$ to the NW of the Orion Trapezium, the
BN/KL region has been, as
the closest region of massive star formation, the subject of extensive studies.
Recently, Rodr\'\i guez et al. (2005) and G\'omez et al. (2005)
reported large proper motions (equivalent to velocities of the order of
a few tens of km s$^{-1}$) for the radio sources associated with the infrared sources
BN and n, as well as for the radio source I. All three objects
are located at the core of the BN/KL region and appear 
to be moving away from a common point where they must all have been 
located about 500 years ago.
Even when these proper motions are now available, there is no
radial velocity information for these three sources, with the
exception of the near-infrared spectroscopic study of BN
made by Scoville et al. (1983), that report an LSR radial
velocity of +21 km s$^{-1}$ for this source.
In this paper we present 7 mm continuum and H53$\alpha$
radio recombination line observations of the BN/KL region in an
attempt to obtain additional information on the radial velocities of
these sources.

\section{Observations}


The 7 mm observations were made in the B configuration
of the VLA of the NRAO\footnote{The National Radio 
Astronomy Observatory is operated by Associated Universities 
Inc. under cooperative agreement with the National Science Foundation.},
during 2007 December 14. The central rest frequency observed was
that of the H53$\alpha$ line, 42951.97 MHz,
and we integrated on-source for a total of
approximately 3 hours. We observed in the spectral line
mode, with 15 channels of 1.56 MHz each (10.9 km s$^{-1}$)
and both circular polarizations. The bandpass calibrator was
0319+415. A continuum channel recorded the
central 75\% of the full spectral window. The absolute amplitude
calibrator was 1331+305 
(with an adopted flux density of 1.47 Jy)
and the phase calibrator was 0541$-$056 (with a bootstrapped flux density
of 1.78$\pm$0.08 Jy). The phase noise rms was about 30$^\circ$,
indicating good weather conditions. The phase center of these observations was at
$\alpha(2000) = 05^h~35^m~14\rlap.^s13;~\delta(2000) = -05^\circ~22{'}~26\rlap.^{''}6$.

The data were acquired and reduced using the recommended VLA procedures
for high frequency data, including the fast-switching mode with a
cycle of 120 seconds. 
Clean maps were
obtained using the task IMAGR of AIPS with the ROBUST parameter set to 0.





\section{Continuum Analysis}

\subsection{Spectral Indices}

In Figure 1 we show the image obtained from the continuum channel.
Three sources, BN, I and n, are evident in the image. No other sources
were detected above a 5-$\sigma$ lower limit of 1.75 mJy in our $1'$
field of view.  The positions, flux
densities, and deconvolved angular sizes of these sources are
given in Table 1. The continuum flux density of the sources
has been obtained from the line-free channels.
The line emission will be discussed below.
The flux density obtained at 7 mm by us 
for BN is in good agreement with the values previously reported in
the literature: 
we obtain a flux density of 28.6$\pm$0.6 mJy, while
values of 31$\pm$5 and 28.0$\pm$0.6 were obtained by Menten \& Reid (1995)
and Chandler \& Wood (1997), respectively. 
In the case of source I, the agreement is acceptable,
since we obtain a flux density of 14.5$\pm$0.7 mJy,
while values of 
13$\pm$2 and 10.8$\pm$0.6 mJy were reported by Menten \& Reid (1995)
and Chandler \& Wood (1997), respectively. 
Careful monitoring would be required
to test if the radio continuum from source I is variable in time. 

The spectral indices determined from our 7 mm observations and the
3.6 cm observations of G\'omez et al. (2008) are given in the last column of Table 2.
Our spectral indices for BN and
I are in excellent agreement in this spectral range with the more detailed analysis 
presented by Plambeck et al. (1995) and Beuther et al. (2004).

We have detected source n for the first time
at 7 mm and this detection allows the first estimate of the spectral index of this source
over a wide frequency range.
The value of 0.2$\pm$0.1 suggests marginally thick free-free emission, as expected in
an ionized outflow. This supports the interpretation of this source
as an ionized outflow by G\'omez et al. (2008).
The position given by us in Table 1 is consistent with the
extrapolation of the proper motions of this source discussed by G\'omez et al. (2008).

\subsection{Deconvolved Angular Sizes}

The radio source I has parameters
consistent with an optically thick free-free source (spectral
index of $1.5\pm0.1$).
Beuther et al. (2004) suggest that this spectral index is either the result of 
optically thick free-free plus dust emission, or $H^-$ free-free emission 
that gives rise to a power-law spectrum with an index of $\sim$1.6. 

In the case of the radio source associated with the infrared source n
we only have an upper limit to its size at 7 mm. In addition,
G\'omez et al. (2008) report important morphological variations
over time in this source
that suggest that comparisons at different frequencies should be made
only from simultaneous observations.

In the case of BN, 
the frequency dependences of flux density and angular size (this last
parameter taken to
be the geometric mean of the major and minor axes reported in Tables 1 and 2) can be accounted for with
a simple model of a sphere of ionized gas in which 
the electron density
decreases as a power-law function of radius, $n_e \propto r^{-\alpha}$. 
In this case, the flux density of the source is expected to go with
frequency as $S_\nu \propto \nu^{(6.2-4\alpha)/(1-2\alpha)}$ and the angular size is expected to go with
frequency as $\theta_\nu \propto \nu^{2.1/(1-2\alpha)}$ (Reynolds 1986).
The frequency dependences of flux density ($S_\nu \propto \nu^{1.1\pm0.1}$) and angular 
size ($\theta_\nu \propto \nu^{-0.36\pm0.12}$) for
BN are consistent with a steeply declining electron density
distribution 
with power law index of
$\alpha = 3.0\pm0.3$. The continuum spectrum of BN produced 
by Plambeck et al. (1995) indicates that a constant
spectral index extends from 5 to 100 GHz.

\section{Analysis of the H53$\alpha$ Recombination Line Emission}

\subsection{Radial LSR Velocity}

We clearly detected the H53$\alpha$ line emission only from BN.
The spectrum is shown in Figure 2. The parameters of
the Gaussian least squares fit to the profile are given in Table 3.
We note that the radial LSR velocity determined by us, $+20.1\pm2.1$
km s$^{-1}$, agrees well with the value of $+21$ km s$^{-1}$
reported by Scoville et al. (1983) from near-IR spectroscopy.
In a single dish study of the H41$\alpha$ line made with an
angular resolution of 24$''$ toward
Orion IRc2, Jaffe \& Mart\'\i n-Pintado (1999) report emission
with $v_{LSR}$ = -3.6 km s$^{-1}$. 
Most likely, this is emission from the ambient H~II region, since
its radial velocity practically coincides with the
value determined for the large H~II region (Orion A) ionized by
the Trapezium stars (e. g. Peimbert et al. 1988).
The single dish observations of the H51$\alpha$ emission
of Hasegawa \& Akabane (1984), made with an angular resolution of 33$''$, 
most probably come also from the ambient ionized gas and not
from BN.

\subsection{LTE Interpretation}

If we assume that the line emission is optically thin and in LTE,
the electron temperature, $T_e^*$, is given by 
(Mezger \& H\"oglund 1967; Gordon 1969; Quireza et al. 2006): 

\begin{equation}\Biggl[{{T_e^*} \over {K}}\Biggr] = \Biggl[7100 \biggl({{\nu_L} \over {GHz}} \biggr)^{1.1}
\biggl({{S_C} \over {S_L}} \biggr) \biggl({{\Delta v} \over {km~s^{-1}}}\biggr)^{-1}
(1 + y^+)^{-1} \Biggr]^{0.87}, \end{equation} 

\noindent where $\nu_L$ is the line frequency, $S_C$ is the continuum flux density,
$S_L$ is the peak line flux density, $\Delta v$ is the FWHM line width, and
$y^+$ is the ionized helium to ionized hydrogen abundance ratio.
In the case of BN, we can adopt $y^+ \simeq 0$ given that the
source is not of very high luminosity, and using the values given in Tables 1 and 3,
we obtain $T_e^* \simeq 8,200$ K. This value is similar to that
determined for the nearby Orion A from radio recombination lines (e. g. Lichten, Rodr\'\i guez, \&
Chaisson 1979). 

It is somewhat
surprising that we get a very reasonable estimate for $T_e^*$ when our previous discussion
seemed to imply that BN is partially optically thick at 7 mm.
One possibility is that we have two effects fortuitously canceling each other. For example, the
optical thickness of the source will diminish the 
line emission, while maser effects (such as those observed
in MWC 349; Mart\'\i n-Pintado et al. 1989) will amplify the line.
However, in an attempt to understand this result in LTE conditions, we will discuss the expected
LTE radio recombination line emission from 
a sphere of ionized gas in which the electron density
decreases as a power-law function of radius, $n_e \propto r^{-\alpha}$. 
As noted before, the  modeling of the continuum emission from such a source 
was presented in detail by Panagia \& Felli (1975) and Reynolds (1986). The radio recombination line emission
for the case $\alpha = 2$ has been discussed by Altenhoff, Strittmatter, \&
Wendker (1981) and Rodr\'\i guez (1982).
Here we generalize the derivation of the recombination line emission 
to the case of $\alpha > 1.5$. This lower limit is
adopted to avoid the total emission from the source to diverge.  

For a sphere of ionized gas, the free-free continuum emission will be given by
(Panagia \& Felli 1975):

\begin{equation}S_C = 2 \pi {{r_0^2} \over {d^2}} B_\nu \int_0^\infty 
\biggl(1 - exp[-\tau_C(\xi)]\biggr)~ \xi~ d\xi, \end{equation} 

\noindent where $r_0$ is a reference radius, $d$ is the distance to the source,
$B_\nu$ is Planck's function, $\xi$ is the projected radius in units of $r_0$,
and $\tau_C(\xi)$ is the continuum optical depth along the line of sight with
projected radius $\xi$. On the other hand, the free-free continuum plus 
radio recombination line emission will be given by an equation similar to eqn. (2), but with the
continuum opacity substituted by the continuum plus line opacity (Rodr\'\i guez 1982):

\begin{equation}S_{L+C} = 2 \pi {{r_0^2} \over {d^2}} B_\nu \int_0^\infty \biggl(1 - exp[-\tau_{L+C}(\xi)]
\biggr) \xi d\xi, \end{equation} 

\noindent where $\tau_{L+C}(\xi)$ is the line plus continuum optical depth along the line of sight with
projected radius $\xi$.

The line-to-continuum ratio will be given by:

\begin{equation}{{S_L} \over {S_C}} = {{S_{L+C} - S_C} \over {S_C}}. \end{equation} 

The opacity of these emission processes depends on projected radius as (Panagia \& Felli 1975):

\begin{equation}\tau(\xi) \propto \xi^{-(2 \alpha -1)}. \end{equation} 

We now introduce the definite integral (Gradshteyn \& Ryzhik 1994)

\begin{equation}\int_0^\infty [1- exp(-\mu x^{-p})]~x~ dx = 
- {{1} \over {p}}~ \mu^{{2} \over{p}}~ \Gamma(-{{2} \over{p}}), \end{equation} 

\noindent valid for $\mu > 0$ and $p > 0$ and with $\Gamma$ being the Gamma function.
Substituting eqns. (2) and (3) in eqn. (4), and using the integral
defined in eqn. (7), it can be shown that

\begin{equation}{{S_L} \over {S_C}} = \Biggl[{{\kappa_L + \kappa_C} 
\over {\kappa_C}} \Biggr]^{1/(\alpha -0.5)} - 1, \end{equation} 

\noindent where $\kappa_L$ and $\kappa_C$ are the line and continuum absorption coefficients
at the frequency of observation, respectively.
In this last step we have also
assumed that the opacity of the line and continuum processes are proportional to
the line and continuum absorption coefficients, respectively, that is, that the
physical depths producing the line and continuum emissions are the
same. Under the LTE assumption, we have
that

\begin{equation}{{\kappa_L} \over {\kappa_C}} = 7100 \biggl({{\nu_L} \over {GHz}} \biggr)^{1.1}
\biggl({{T_e^*} \over {K}} \biggr)^{-1.1} \biggl({{\Delta v} \over {km~s^{-1}}}\biggr)^{-1}
(1 + y^+)^{-1}. \end{equation} 

For $\nu \leq$ 43 GHz and typical parameters of an H II region, we
can see from eqn. (8) that $\kappa_L<\kappa_C$, and
eqn. (7) can be approximated by:

\begin{equation}{{S_L} \over {S_C}} \simeq {{1} \over 
{(\alpha -0.5)}} \Biggl[{{\kappa_L} \over {\kappa_C}} \Biggr]. \end{equation} 

That is, the expected optically-thin, LTE line-to-continuum ratio:

\begin{equation}{{S_L} \over {S_C}} \simeq \Biggl[{{\kappa_L} \over {\kappa_C}} \Biggr], \end{equation} 

\noindent becomes attenuated by a factor $1/(\alpha -0.5)$. In the case of $\alpha = 2$,
the factor is 2/3, and we reproduce the result of Altenhoff, Strittmatter, \&
Wendker (1981) and Rodr\'\i guez (1982). In the case of BN, we have that $\alpha \simeq 3$, and
we expect the attenuation factor to be 2/5. If BN can be modeled this way, we would have expected
to derive electron temperatures under the LTE assumption (see eqn. 1) of order   

\begin{equation}T_e^*(\alpha = 3) \simeq 2.2~ T_e^*(thin). \end{equation} 

However, from the discussion in the first paragraph of this section observationally
we determine that 

\begin{equation}T_e^*(\alpha = 3) \simeq T_e^*(thin). \end{equation} 

Summarizing: i) BN seems to have significant optical depth in the continuum at
7 mm, ii) this significant optical depth should attenuate the observed recombination
line emission with respect to the optically-thin case, but iii) the line emission seems
to be as strong as in the optically-thin case. 

As possible explanations for the ``normal'' (apparently optically-thin and in LTE)
radio recombination line emission
observed from BN we can think of two options.
The first is that, as noted before, there is a non-LTE line-amplifying
mechanism that approximately compensates for the optical depth attenuation.
The second possibility is that the free-free emission from BN at 7 mm is already optically thin.
However, this last possibility seems to be in contradiction with the results
of Plambeck et al. (1995) that suggest a single spectral index 
from 5 to 100 GHz. Observations of radio recombination lines around
100 GHz are needed to solve this problem.

A comparison with the H53$\alpha$ emission from the hypercompact H~II
region G28.20-0.04N is also of interest.  
The continuum flux densities from this source at 
21, 6, 3.6, and 2 cm are 49, 135, 297, and 543 mJy, respectively
(Sewilo et al. 2004). At 7 mm the continuum flux density is 641 mJy
(Sewilo et al. 2008), indicating
that the source has become optically thin at this wavelength.
Using the H53$\alpha$ line parameters given by (Sewilo et al. 2008)
we derive an LTE electron temperature of $T_e^* \simeq 7,600$ K, 
similar to the value for BN and in this case consistent with
the optically-thin nature of G28.20-0.04N.  

The non detection of H53$\alpha$ emission from radio source I is consistent
with its expected large optical depth. The formulation above implies $\alpha \simeq 5$, and an
attenuation factor of 2/9. 
This confirms the notion that BN and radio source I are two sources
intrinsically very different in nature.
This difference is also evident in the brightness temperature of both sources.
At 7 mm, the brightness temperature of a source is

\begin{equation}\Biggl[{{T_B} \over {K}} \Biggr] \simeq 0.96 \Biggl[{{S_\nu} \over {mJy}} 
\Biggr] \Biggl[{{\theta_{maj} \times
\theta_{min}} \over {arcsec^2}} \Biggr]^{-2}. \end{equation} 

Using the values of Table 1, we get $T_B \simeq$ 7,800 K for BN, confirming
its nature as photoionized gas. However, for the radio source I we get
$T_B \simeq$ 2,600 K. So, even when source I seems to be optically thick, its
brightness temperature is substantially lower than that expected for
a photoionized region. Reid et al. (2007) have discussed as possible
explanations for this low brightness temperature $H^-$ free-free opacity or 
a photoionized disk.

Following the discussion of Reid et al. (2007), we consider
it unlikely that dust emission could be a dominant contributor to the 7 mm emission of BN or
Orion I. A dense, warm, dusty disk would be expected to show many molecular lines at
millimeter/submillimeter wavelengths. While Beuther et al. (2006) and Friedel
\& Snyder(2008) find numerous, strong,
molecular lines toward the nearby "hot core", they find no strong lines toward the position of
Orion I (with the exception of
the strong SiO masers slightly offset from Orion I) or BN.
Also, the brightness temperatures derived by us at 7 mm (7,800 K for BN and
2,600 K for source I) are 
high enough to sublimate dust and suggest that free-free emission from
ionized gas dominates the continuum emission.
Finally, the continuum spectra of BN and of source I measured by Plambeck et al.(1995)
and Beuther et al. (2006), respectively, suggest that the dust
emission becomes dominant only above $\sim$300 GHz.

In the case of source n, no detection was expected given its
weakness even in the continuum.

\subsection{Spatial Distribution of the H53$\alpha$ Line Emission}

The H53$\alpha$ line emission in the individual velocity
channels shows evidence of structure but unfortunately the signal-to-noise
ratio is not large enough to reach reliable conclusions from the
analysis of these individual channels. However, an image
with good signal-to-noise ratio can be obtained averaging over the velocity
range of -21.2 to +66.1 km s$^{-1}$, using the task MOMNT in
AIPS. This line image is compared
in Figure 3 with a continuum image
made from the line-free channels.
The larger apparent size of the continuum image is simply the
result of its much better signal-to-noise ratio.
For the total line emission we obtain an upper limit of
$0\rlap.{''}12$ for its size, that is consistent with the
size of the continuum emission given in Table 1.
We also show images of the blueshifted (-21.2 to +22.5 km s$^{-1}$)
and redshifted (+22.5 to 66.1 km s$^{-1}$) line emission in Figure 3.
The cross in the figure indicates the centroid of the total line
emission. The centroid of the line emission does not appear to
coincide with the centroid of the continuum emission and
we attribute this to opacity effects.

An interesting conclusion comes from comparing the total
line emission, with the blueshifted and redshifted components.
The blueshifted emission seems slightly shifted to the SW, while the
redshifted emission seems slightly shifted to the NE, suggesting a
velocity gradient. This result supports the suggestion of
Jiang et al. (2005) of the presence of an outflow in BN along a
position angle of 36$^\circ$. Given the modest signal-to-noise ratio
of the data, it is difficult to estimate the magnitude
of the velocity shift and we crudely assume it is of order one
channel ($\sim$10 km s$^{-1}$), since most of the line
emission is concentrated in the central two channels
of the spectrum (see Figure 2). The position shift between the blueshifted and
the redshifted emissions is $0\rlap.{''}028 \pm 0\rlap.{''}007$
($12 \pm 3$ AU at the distance of 414 pc given by Menten et al. 2007), significant to the 
4-$\sigma$ level. Unfortunately, the data of Jiang et al. (2005) does not
include line observations and there is no kinematic information in their paper to
compare with our results.

The small velocity gradient observed by us in BN is consistent with a
slow bipolar outflow but also with Keplerian rotation around a central mass
of only 0.2 $M_\odot$. 
 
\section{Conclusions}

We presented observations of the H53$\alpha$ recombination line
and adjacent continuum toward the Orion BN/KL region.
In the continuum we detect the BN object, the radio source 
I (GMR I) and the radio counterpart of the infrared source n 
(Orion-n) and discuss its parameters. 
In the H53$\alpha$ line we only detect the BN object,
the first time that radio recombination lines have been detected from this source.
The LSR radial velocity of BN from the H53$\alpha$ line, $v_{LSR} = 20.1 \pm 2.1$
km s$^{-1}$,
is consistent with that found from previous studies in near-infared lines,
$v_{LSR} = 21$ km s$^{-1}$.
We discuss the line-to-continuum ratio from BN and present evidence
for a possible velocity gradient across this source. 

\acknowledgments

LFR and LAZ acknowledge the support
of CONACyT, M\'exico and DGAPA, UNAM.



{\it Facilities:} \facility{VLA}

\clearpage

\clearpage

\begin{table}
\begin{center}
\small
\caption{Parameters of the 7 mm Continuum Sources in the Orion BN/KL Region\label{tbl-1}}
\begin{tabular}{lcccc}
\tableline\tableline
 &\multicolumn{2}{c}{Position$^a$} & Total Flux
& \\
\cline{2-3}
Source &  $\alpha$(J2000) & $\delta$(J2000) & Density (mJy) &
Deconvolved Angular Size$^b$  \\
\tableline
BN & 05 35 14.110 & -05 22 22.73 & 28.6$\pm$0.6
& $0\rlap.{''}07 \pm 0\rlap.{''}01 \times 0\rlap.{''}05 \pm 0\rlap.{''}01;~ +45^\circ
\pm 29^\circ$  \\
n & 05 35 14.359 &  -05 22 32.78 & 3.0$\pm$0.6
& $\leq 0\rlap.{''}2$ \\
I & 05 35 14.514 & -05 22 30.57 & 14.5$\pm$0.7
& $0\rlap.{''}09 \pm 0\rlap.{''}01 \times 0\rlap.{''}06 \pm 0\rlap.{''}02;~ +134^\circ
\pm 44^\circ$  \\
\tableline
\end{tabular}
\tablenotetext{a}{Units of right
ascension are hours, minutes, and seconds
and units of declination are degrees, arcminutes, and arcseconds. Positional accuracy
is estimated to be $0\rlap.{''}01$.}
\tablenotetext{b}{Major axis $\times$ minor axis; position angle of major axis.}
\end{center}
\end{table}

\clearpage

\begin{table}
\begin{center}
\small
\caption{Parameters of the 3.6 cm Continuum Sources in the Orion BN/KL Region\label{tbl-2}}
\begin{tabular}{lccccc}
\tableline\tableline
 &\multicolumn{2}{c}{Position$^a$} & Total Flux
& & Spectral \\
\cline{2-3}
Source &  $\alpha$(J2000) & $\delta$(J2000) & Density (mJy) &
Deconvolved Angular Size$^b$ & Index \\
\tableline
BN & 05 35 14.110 & -05 22 22.74 & 4.8$\pm$0.1
& $0\rlap.{''}17 \pm 0\rlap.{''}01 \times 0\rlap.{''}08 \pm 0\rlap.{''}02;~ +63^\circ
\pm 7^\circ$ & 1.1$\pm$0.1  \\
n & 05 35 14.355 &  -05 22 32.78 & 2.2$\pm$0.2
& $0\rlap.{''}50 \pm 0\rlap.{''}04 \times \leq 0\rlap.{''}09;~ +20^\circ
\pm 3^\circ$ & 0.2$\pm$0.1  \\
I & 05 35 14.514 & -05 22 30.56 & 1.2$\pm$0.1
& $0\rlap.{''}19 \pm 0\rlap.{''}04 \times \leq 0\rlap.{''}15;~ +136^\circ
\pm 23^\circ$ & 1.5$\pm$0.1  \\
\tableline
\end{tabular}
\tablenotetext{a}{Units of right
ascension are hours, minutes, and seconds
and units of declination are degrees, arcminutes, and arcseconds. Positional accuracy
is estimated to be $0\rlap.{''}01$.}
\tablenotetext{b}{Major axis $\times$ minor axis; position angle of major axis.}
\end{center}
\end{table}

\clearpage

\begin{table}
\begin{center}
\small
\caption{Parameters of H53$\alpha$ Recombination Line from the BN Object\label{tbl-3}}
\begin{tabular}{lcc}
\tableline\tableline
Peak Flux &  Half Maximum  & LSR Radial \\
Density (mJy) & Line Width (km s$^{-1}$) & Velocity (km s$^{-1}$) \\
\tableline
10.4$\pm$1.1 & 39.0$\pm$4.9 & 20.1$\pm$2.1 \\
\tableline
\end{tabular}
\end{center}
\end{table}

\clearpage

\begin{figure}
\epsscale{.80}
\plotone{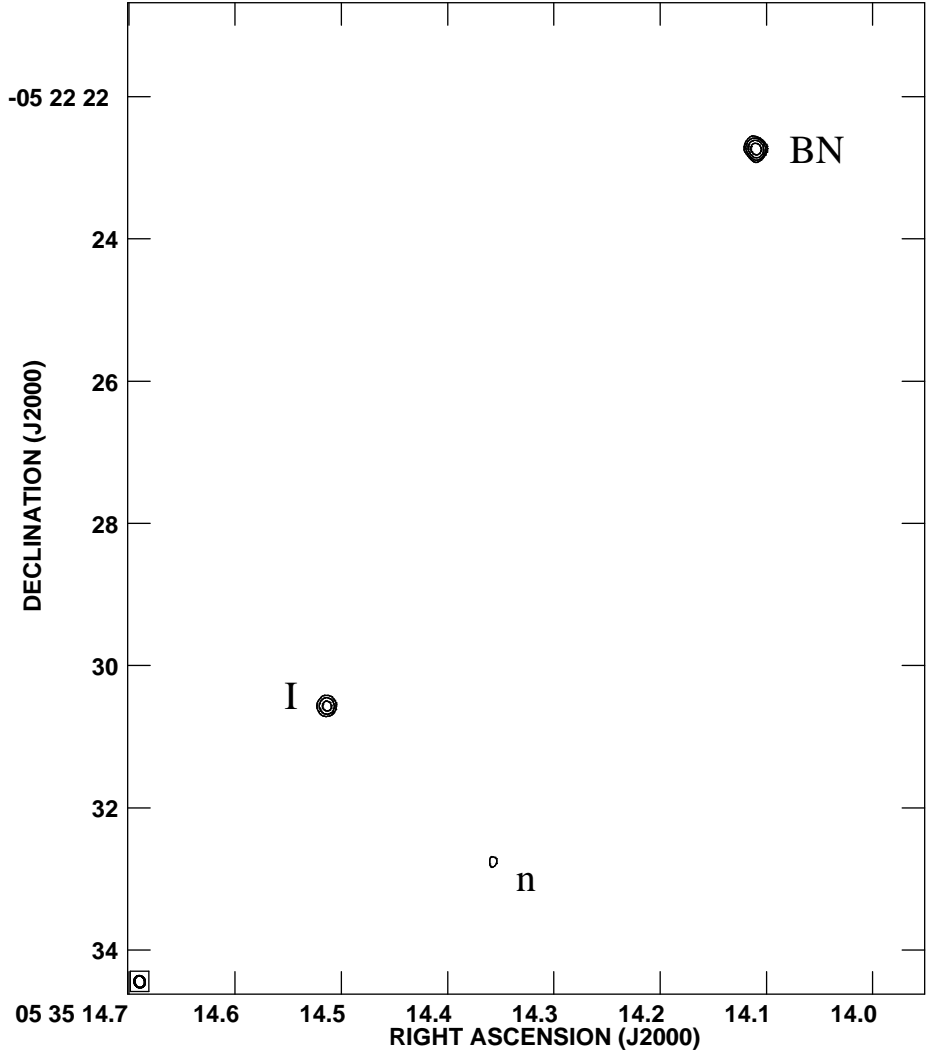}
\caption{Contour image of the 7 mm continuum emission
from the BN/KL region. The three sources detected are marked
with their names.
Contours are
-5, 5, 10, 20, and 40 times 0.35 mJy beam$^{-1}$, the rms noise of the image. 
The half power contour of the synthesized beam 
($0\rlap.{''}17 \times 0\rlap.{''}15$
with a position angle of $+17^\circ$) is shown in the bottom left corner.
\label{fig1}}
\end{figure}

\clearpage

\begin{figure}
\epsscale{.80}
\plotone{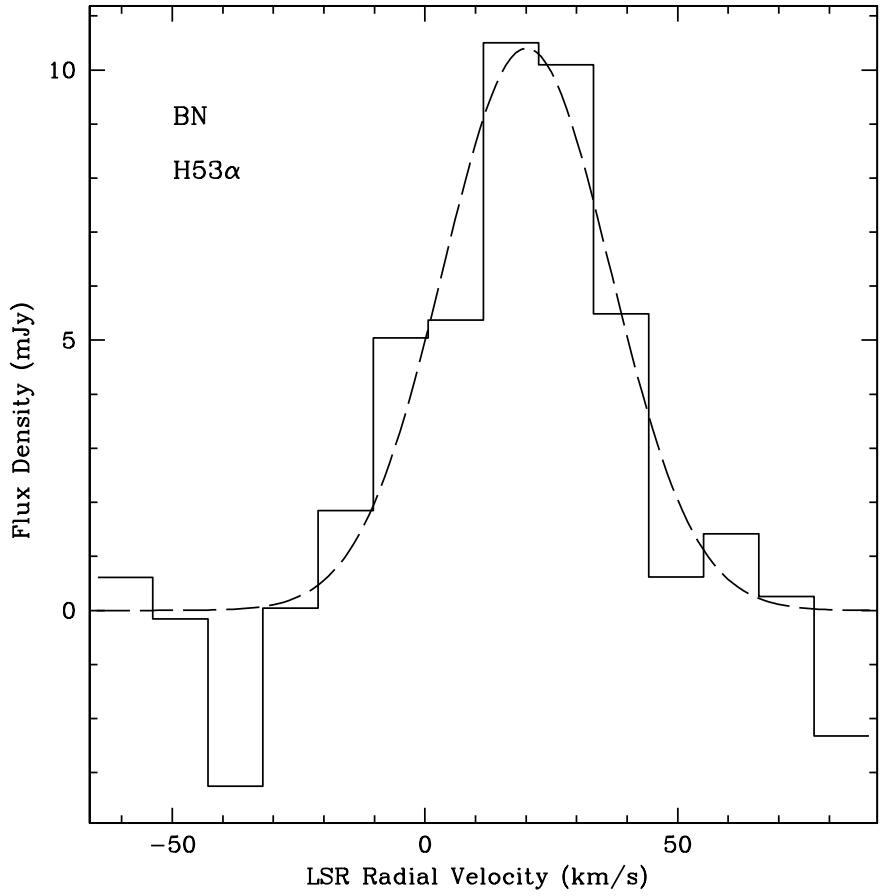}
\caption{Spectrum of the H53$\alpha$ line emission
from BN. The dashed line is the least squares fit
to the data, whose parameters are given in Table 3.
\label{fig2}}
\end{figure}

\begin{figure}
\epsscale{1.0}
\plotone{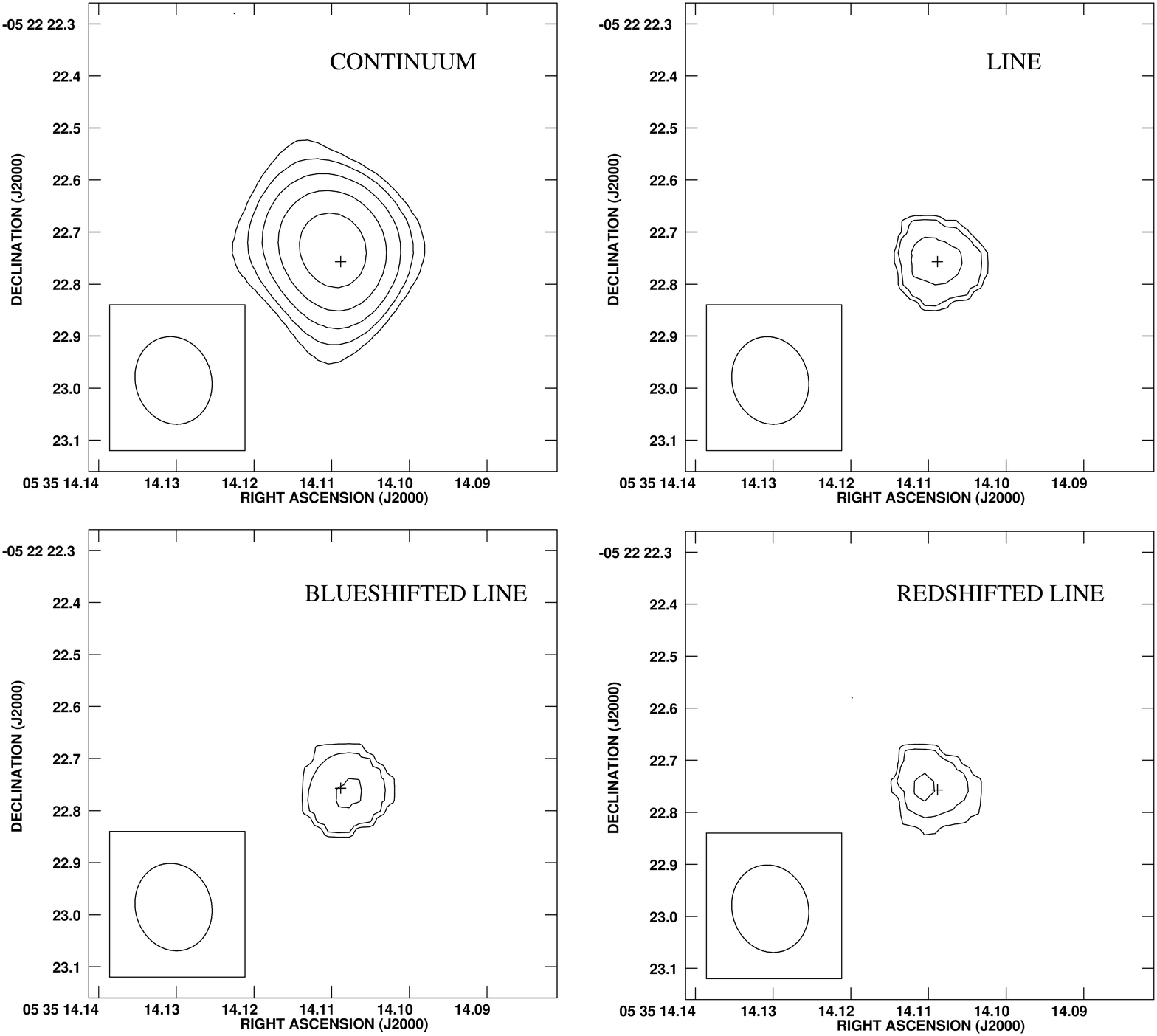}
\caption{Contour images of the 7 mm continuum emission
(top left), the average of the total (-21.2 to +66.1 km s$^{-1}$)
H53$\alpha$ line emission (top right),
the average of the blueshifted (-21.2 to +22.5 km s$^{-1}$) 
H53$\alpha$ line emission (bottom left),
and the average of the redshifted (+22.5 to 66.1 km s$^{-1}$)
H53$\alpha$ line emission (bottom right) 
from the BN object.
Contours are
1, 2, 4, 8, and 16 times 1.0 mJy beam$^{-1}$ for the continuum image.
Contours are 1, 2, and 4 times
12.0 mJy beam$^{-1}$ km s$^{-1}$ for the total line image
and 16.0 mJy beam$^{-1}$ km s$^{-1}$ for the blueshifted and
redshifted line images.
The cross marks the peak position of the average of the total H53$\alpha$ line emission. 
The half power contour of the synthesized beam ($0\rlap.{''}17 
\times 0\rlap.{''}15$
with a position angle of $+17^\circ$) is shown in the bottom left corner
of the images.
\label{fig3}}
\end{figure}







\begin{thebibliography}{}


\bibitem[Altenhoff et 
al.(1981)]{1981A&A....93...48A} Altenhoff, W.~J., Strittmatter, P.~A., \& Wendker, H.~J.\ 1981, \aap, 93, 48

\bibitem[Beuther et al.(2004)]{2004ApJ...616L..31B} Beuther, H., et al.\ 
2004, \apjl, 616, L31 

\bibitem[Beuther et al.(2006)]{2006ApJ...636..323B} Beuther, H., et al.\
2006, \apj, 636, 323

\bibitem[Chandler 
\& Wood(1997)]{1997MNRAS.287..445C} Chandler, C.~J., \& Wood, D.~O.~S.\ 1997, \mnras, 287, 445 


\bibitem[Friedel 
\& Snyder(2008)]{2008ApJ...672..962F} Friedel, D.~N., \& Snyder, L.~E.\ 2008, \apj, 672, 962 



\bibitem[]{go05} G\'omez, L., Rodr\'{\i}guez, L.F., Loinard, L., 
Poveda, A., Lizano, S., \& Allen, C.\ 
2005, ApJ, 635, 1166

\bibitem[]{go07} G\'omez, L., et al. 2008, ApJ, 685, 333

\bibitem[]{go69} Gordon, M.~A.\ 1969, \apj, 158, 
479 

\bibitem[]{gr94} Gradshteyn, I.~S., \& Ryzhik, I.~M.\ 1994, New York: Academic Press, 
5th ed., edited by Jeffrey, A., p. 386


\bibitem[Hasegawa 
\& Akabane(1984)]{1984ApJ...287L..91H} Hasegawa, T., \& Akabane, K.\ 1984, \apjl, 287, L91 

\bibitem[Jaffe 
\& Mart{\'{\i}}n-Pintado(1999)]{1999ApJ...520..162J} Jaffe, D.~T., \& 
Mart{\'{\i}}n-Pintado, J.\ 1999, \apj, 520, 162 

\bibitem[]{ji05} Jiang, Z., Tamura, M., 
Fukagawa, M., Hough, J., Lucas, P., Suto, H., Ishii, M., 
\& Yang, J.\ 2005, \nat, 437, 112 

\bibitem[Lichten et al.(1979)]{1979ApJ...229..524L} Lichten, S.~M., 
Rodriguez, L.~F., \& Chaisson, E.~J.\ 1979, \apj, 229, 524 

\bibitem[Martin-Pintado et 
al.(1989)]{1989A&A...215L..13M} Mart\'\i n-Pintado, J., 
Bachiller, R., Thum, C., \& Walmsley, M.\ 1989, \aap, 215, L13 


\bibitem[Menten 
\& Reid(1995)]{1995ApJ...445L.157M} Menten, K.~M., \& Reid, M.~J.\ 1995, \apjl, 445, L157 

\bibitem[]{me07} Menten, K. M., Reid, M. J., Forbrich, J., \& Brunthaler, A.
2007, A\&A, 474, 515


\bibitem[Panagia 
\& Felli(1975)]{1975A&A....39....1P} Panagia, N., \& Felli, M.\ 1975, \aap, 39, 1 

\bibitem[Peimbert et al.(1992)]{1992ApJ...395..484P} Peimbert, M., 
Rodriguez, L.~F., Bania, T.~M., Rood, R.~T., 
\& Wilson, T.~L.\ 1992, \apj, 395, 484 

\bibitem[Plambeck et al.(1995)]{1995ApJ...455L.189P} Plambeck, R.~L., 
Wright, M.~C.~H., Mundy, L.~G., \& Looney, L.~W.\ 1995, \apjl, 455, L189 

\bibitem[]{qui06}Quireza, C., Rood, 
R.~T., Bania, T.~M., Balser, D.~S., \& Maciel, W.~J.\ 2006, \apj, 653, 1226


\bibitem[Reid et al.(2007)]{2007ApJ...664..950R} Reid, M.~J., Menten, 
K.~M., Greenhill, L.~J., \& Chandler, C.~J.\ 2007, \apj, 664, 950 

\bibitem[]{re86} Reynolds, S. P. 1986, ApJ, 304, 713

\bibitem[Rodriguez(1982)]{1982RMxAA...5..179R} Rodriguez, L.~F.\ 1982, 
Revista Mexicana de Astronomia y Astrofisica, 5, 179 


\bibitem[]{ro05} Rodr\'{\i}guez, L. F., Poveda, A., Lizano, S., \& Allen, C. 
2005, ApJ, 627, L65

\bibitem[Scoville et al.(1983)]{1983ApJ...275..201S} Scoville, N., 
Kleinmann, S.~G., Hall, D.~N.~B., \& Ridgway, S.~T.\ 1983, \apj, 275, 201 

\bibitem[Sewilo et al.(2004)]{2004ApJ...605..285S} Sewilo, M., Churchwell, 
E., Kurtz, S., Goss, W.~M., \& Hofner, P.\ 2004, \apj, 605, 285 

\bibitem[Sewilo et al.(2008)]{2008arXiv0803.2872S} Sewilo, M., Churchwell, 
E., Kurtz, S., Goss, W.~M., 
\& Hofner, P.\ 2008, 681, 350



\end{thebibliography}
\end{document}